\documentclass[aps,prl,twocolumn,secnumarabic,amssymb,amsmath,10pt]{revtex4-2}
\usepackage{graphicx}
\usepackage{bm}
\usepackage{siunitx}
\usepackage{lipsum} 
\usepackage{amsmath}
\usepackage{mhchem}
\begin{document}

\title{Distinguishing and Separating In-Plane Hall Responses}
\author{Soumya Sankar}
\affiliation{Department of Physics, The Hong Kong University of Science and Technology, Clear Water Bay, Kowloon, Hong Kong SAR}

\author{Xingkai Cheng}
\affiliation{Department of Physics, The Hong Kong University of Science and Technology, Clear Water Bay, Kowloon, Hong Kong SAR}

\author{Junwei Liu}
\email{liuj@ust.hk}
\affiliation{Department of Physics, The Hong Kong University of Science and Technology, Clear Water Bay, Kowloon, Hong Kong SAR}

\author{Berthold J\"ack}
\email{bjaeck@ust.hk}
\affiliation{Department of Physics, The Hong Kong University of Science and Technology, Clear Water Bay, Kowloon, Hong Kong SAR}

\date{\today}

\begin{abstract}

Electric Hall effects generated by an in-plane magnetic field have recently gained attention owing to their intrinsic origin in topological electronic states and potential application in magnetic field sensing. In pratice, the measured transverse electric voltage typically combines contributions from multiple phenomena, such as anisotropy and Berry curvature effects, leading to interpretative ambiguities of the measurement signal. Here, we introduce a universal framework that disentangles these contributions via their distinct field‑reversal symmetries and angular dependencies. Leveraging a 12‑terminal Hall bar for independent control of the electric and in-plane magnetic field directions, we exemplify this method by analyzing the transverse electric voltage recorded on the the ferromagnetic Weyl semimetal Fe$_3$Sn in an in-plane geometry. The standardized approach presented in this work will guide future studies of in-plane Hall responses in magnetic and topological materials.
\end{abstract}

\maketitle

{\em Introduction}.--Topological electronic wavefunctions with finite Berry curvature can give rise to novel electric Hall effects, such as the anomalous Hall (AHE)~\cite{nagaosa2010anomalous,ye2018massive,deng2018gate,pan2022giant,yin2018giant,takagi2025spontaneous} and nonlinear anomalous Hall effect (NLAHE)~\cite{sankar2024experimental,ma2019observation,gao2023quantum,wang2023quantum}. These responses are of great interest for fundamental condensed matter physics and their potential applications in magnetic field sensing~\cite{ni2016ultrahigh}, broadband communication networks~\cite{sankar2025broadband}, and energy harvesting~\cite{onishi2024high}, motivating research into topological and magnetic quantum materials. Conventionally, the electric Hall effect is associated with a transverse voltage generated within the sample $xy$-plane, either by ferromagnetism or by an external magnetic field applied along the out-of-plane $z$-direction.  These well-known anomalous (AHEs) and ordinary Hall (OHEs) effects only require breaking of time-reversal symmetry \(\mathcal{T}\) and are typically isotropic within the $xy$-plane, meaning are independent of specific device configurations, i.e., the directions of current injection and voltage detection.

\begin{figure}
    \centering
    \includegraphics[width=\columnwidth]{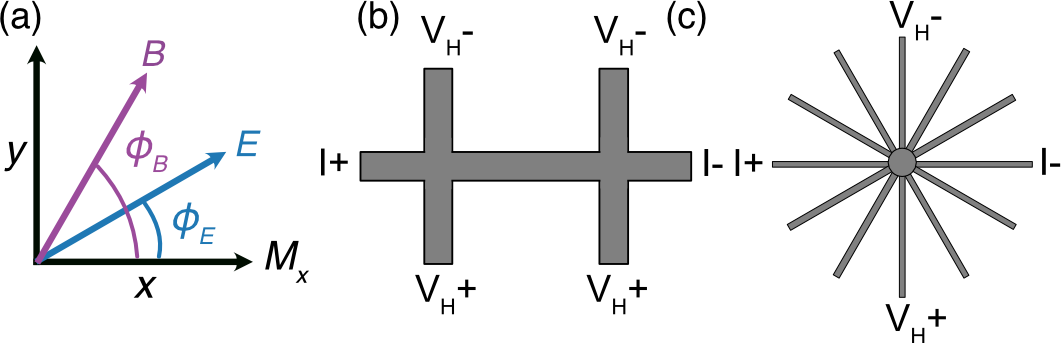}
    \caption{(a) Shown is the in-plane component of the sample magnetization \(M_x\) defined to be parallel to the $+x$ direction. The azimuthal angles $\phi_E$ and $\phi_B$ are defined as the angles between the direction of the electric field $\mathbf{E}$ and the $x$-axis, and the direction of the in-plane magnetic field $\mathbf{B}_\parallel$ and the $x$-axis, respectively. (b) Shown is a schematic of a regular Hall bar geometry with current ($I$) sourced from left to right while measuring the Hall voltage ($V_H$) across the top and bottom contacts, providing access to only four values of azimuthal angle $\phi_E$. (c) Shown is a schematic of a circular Hall bar with 12 legs. Current $I$ is sourced from left to right while the Hall voltage $V_H$ is measured across the top and bottom contacts, providing access to twelve discrete values of the azimuthal angle $\phi_E$.}
    \label{fig:1}
\end{figure}

Recent attention has been drawn to particularly interesting scenarios in which the electric field $\bf E$, the in‑plane magnetic field $\bf B_{\parallel}$, and the sample magnetization $\bf M_{\parallel}$ all lie within the sample plane, giving rise to in‑plane Hall responses~\cite{mcguire2003anisotropic,tang2003giant, nakamura2024observation,sankar2025room,zhou2022heterodimensional,liang2018anomalous}. In such configurations, the three external knobs ($\bf E$, ${\bf B}_{\parallel}$, and ${\bf M}_{\parallel}$) are no longer independent but engage in multiple complex couplings, making the resulting transverse voltage a sum of several coexisting phenomena. The measured signal is thus generally anisotropic within the $xy$-plane and depends intricately on the relative angles among $E$, $B_{\parallel}$, and $M_{\parallel}$. For example, the breaking of the three-fold out-of-plane rotation symmetry ($C_{3z}$) by ${\bf M}_\parallel\neq0$ generally causes the appearance of the symmetric transverse resistance (STR) effect, an anisotropy effect due to intrinsic magnetization~\cite{mcguire2003anisotropic}, which depends on the relative angle $\phi_E$ between $\bf M_\parallel$ and $\mathbf{E}$~\cite{SI}, as defined in Fig.~\ref{fig:1}(a). Moreover, the presence of $\mathbf{B_\parallel}$ can cause the planar Hall effect (PHE)~\cite{tang2003giant, bowen2005order} through magnetic field-induced anisotropic magnetoresistance~\cite{zhong2023recent}, which depends on the relative angle $\phi_B-\phi_E$ between $\mathbf{B_\parallel}$ and $\mathbf{E}$. Because both STR and PHE are anisotropy effects, they are even under magnetization/magnetic field reversal. 

On the other hand, when both $\mathcal{T}$ and specific crystalline symmetries, such as out-of-plane mirror and rotation symmetries, are broken by $\bf M_\parallel$, the anomalous in-plane Hall effect (AIPHE) can be detected under the application of ${\bf B}_\parallel$~\cite{nakamura2024observation,sankar2025room,zhou2022heterodimensional,liang2018anomalous}. This effect depends only on the relative angle $\phi_B$ between $\bf M_\parallel$ and $\mathbf{B_\parallel}$ [Fig.~\ref{fig:1}(a)] and is odd under $\phi_B\rightarrow\phi_B+\pi$. The AIPHE has attracted considerable recent attention because it is intimately connected to the intrinsic quantum properties of the electronic band structure such as the Berry curvature, quantum metric, and the chirality of Weyl points~\cite{mcguire2003anisotropic,SI,liu2013plane,sun2022possible, sankar2025room}. In particular, the presence of Weyl points, which act as monopoles of Berry curvature in momentum space, can give rise to a large AIPHE signal which can be effectively tuned by the in-plane magnetic field~\cite{sankar2025room}.

As a consequence, these different phenomena (STR, PHE, and IPHE) jointly contribute to the measured transverse voltage signal $V_{xy}$ under commonly occurring experimental conditions. Yet, separating these effects has remained challenging for two primary reasons. First, commonly used device geometries, such as the van der Pauw and conventional Hall bar [Fig.~\ref{fig:1}(b)] geometries used in the study of bulk crystals and thin film devices, respectively, do not offer the required angle resolution with respect to $\phi_E$. Second, a universal symmetry‑based framework to disentangle these contributions has not been developed and validated. This causes interpretative ambiguities of $V_{xy}$, limiting the use of in-plane Hall measurements for the study of topological quantum materials and the associated transport phenomena.

In this Letter, we overcome these limitation by introducing a symmetry‑guided experimental protocol that simultaneously achieves two goals: (i) it unambiguously disentangles the PHE, STR, and AIPHE contributions using their different field‑reversal symmetries and angular dependencies; (ii) it employs the direction of the in‑plane magnetic field as a clean, continuously tunable parameter to modulate the AIPHE signal, a true Hall effect. This paradigm opens new opportunities for exploiting in‑plane Hall effects in functional devices, including magnetic field sensors, spin‑orbitronics, and energy‑efficient signal processing.

\vspace{0.25cm}
{\em Theoretical Background}.--The transverse  linear current response \( J_{\mu} \) to an electric field \( E_{\nu} \) applied along the longitudinal direction is governed by the conductivity tensor \( \sigma_{\mu\nu} \). For simplicity, we define the longitudinal and transverse directions being along the $x-$ and $y-$direction, respectively, such that \( J_{y} = \sigma_{yx} E_{x} \). In the case of dissipation-less transport, the antisymmetric component of this tensor, \( \sigma_{yx}^{\text{A}} \), is of particular interest. It is intimately connected to the Berry curvature of the occupied Bloch states and underpins the intrinsic AHEs~\cite{nagaosa2010ahe}. For practical purposes, it is often instructive to consider the transverse resistivity $\rho_{yx}\approx\sigma_{yx}^{-1}$, because it is directly proportional to $V_{yx}$.

In a realistic experimental setting, $V_{yx}$ measured under the application of an in-plane magnetic field ${\bf B}_\parallel$ can comprise contributions from various phenomena, such as the STR, the PHE, and the AIPHE. As we discuss in the following, each of these phenomena is characterized by a specific dependence on $\phi_B$ and $\phi_E$. It is this unique dependence on $\phi_B$ and $\phi_E$ by which the contribution of these phenomena to $\rho_{\rm yx}$ can be qualitatively distinguished.

\begin{figure}
    \centering
    \includegraphics[width=\columnwidth]{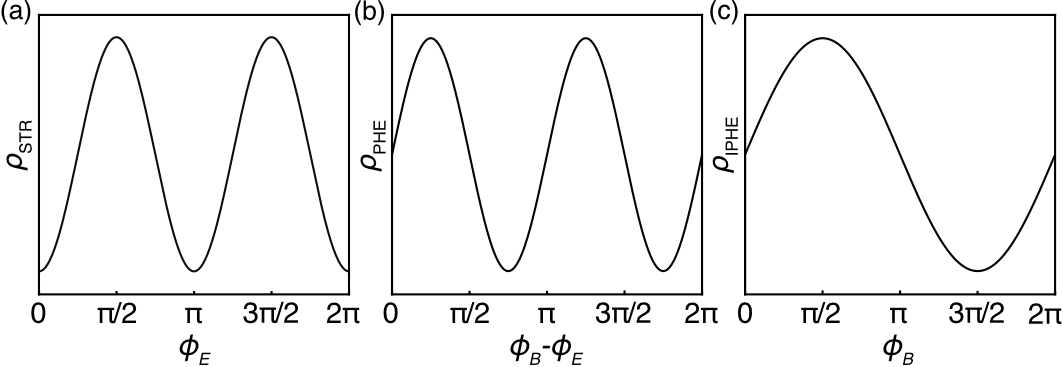}
    \caption{(a) Schematic illustration of the  angular dependence of the symmetric transverse resistivity ($\rho_\text{STR}$) as a function of azimuthal angle $\phi_E$ with a periodicity of $\pi$. (b) Shown is the angular dependence of the planar Hall effect resistivity ($\rho_\mathrm{PHE}$) as a function of the azimuthal angle difference $\phi_B - \phi_E$ with a periodicity of $\pi$. (c) Shown is the angular dependence of the anomalous in-plane Hall resistivity ($\rho_\mathrm{AIPHE}$) as a function of azimuthal angle $\phi_B$ with a periodicity of $2\pi$.}
    \label{fig:2}
\end{figure}

To this end, it is instructive to decompose $\rho_{\rm yx}$ into magnetic field-symmetric $\rho_{\rm sym}(\phi_B,\phi_E)$ and asymmetric $\rho_{\rm asym}(\phi_B,\phi_E)$ parts:
\begin{equation}
\rho_{\rm yx}(\phi_B, \phi_E) =\rho_{\mathrm{sym}}(\phi_B, \phi_E) + \rho_{\mathrm{asym}}(\phi_B, \phi_E).
\end{equation}
We first consider the field-symmetric part $\rho_{\mathrm{sym}}(\phi_B, \phi_E)$, which in general can contain the sum of two contributions, the STR and the PHE:
\begin{equation}
\rho_{\mathrm{sym}}(\phi_B, \phi_E) = \rho_{\mathrm{STR}}(\phi_E) + \rho_{\mathrm{PHE}}(\phi_B - \phi_E).
\end{equation}
The STR can appear in materials with finite ${\bf M}_\parallel$, which breaks the crystalline $C_{3z}$ rotation symmetry. Hence, the STR effect is a magnetic anisotropy effect and can appear in the absence of an externally applied magnetic field~\cite{mcguire2003anisotropic}. Its spatial anisotropy thus solely depends on the azimuthal angle $\phi_E$ via a sinusoidal dependence, $\rho_{\mathrm{STR}}(\phi_E)\propto\sin(2\phi_E)$~\cite{SI}, as illustrated in Fig.~\ref{fig:2}(a), given $\bf M$ remains pinned along the preferred easy-plane direction when the external magnetic field is weak compared to the magnetocrystalline anisotropy energy.

The PHE, on the other hand, can appear when both $\mathbf{E}$ and $\mathbf{B}_{\parallel} =B(\cos\phi_B,\sin\phi_B,0)$ are applied in the sample plane. The PHE originates from magnetic-field induced magnetic anisotropy~\cite{liu2019nontopological,seemann2011origin}. Its contribution to the symmetric part of the diagonal resistivity tensor elements thus depends on the relative angle between the magnetic and electric field, $\rho_{\mathrm{PHE}}\propto\sin(2\phi_B - 2\phi_E)$~\cite{SI}, as illustrated in Fig.~\ref{fig:2}(b). The $\pi$-periodicity of $\rho_{\mathrm{PHE}}$ with respect to $\phi_B$ reflects the even character under magnetic field reversal.

Next, we consider the antisymmetric part $\rho_{\text{asym}}(\phi_B)$ of $\rho_{\text{yx}}$ that can generally contain contributions from the AHE, OHE, and AIPHE. Because only the AIPHE has a distinct dependence on $\phi_B$ (see Ref.~\cite{SI} for discussion), we here focus on the contribution of the AIPHE and set $\rho_{\mathrm{asym}}(\phi_B)=\rho_{\mathrm{AIPHE}}(\phi_B)$. The AIPHE can arise from exchange coupling between ${\bf B}_\parallel$ or ${\bf M}_\parallel$ and the spin of conduction electrons leading to a modulation of the Hall resistance by an in-plane magnetic field~\cite{liu2013plane, sun2022possible, miao2024engineering, sankar2025room}, and thus only depends on $\phi_B$. Distinct from the STR and PHE, the AIPHE is odd under field reversal, $\phi_B\rightarrow\phi_B+\pi$. Hence, its contribution to $\rho_{\mathrm{asym}}$ can be described by a sinusoidal dependence $\rho_{\mathrm{IPHE}} \propto\cos(\phi_B)$ with $2\pi$-periodicity~\cite{SI}, as illustrated in Fig.~\ref{fig:2}(c). 

In the following, we use this distinct dependence of STR, PHE, and IPHE on $\phi_B$ and $\phi_E$ to separate their contributions to the experimentally measured transverse voltage by using circular Hall bar devices.

\vspace{0.25cm}
{\em Results}.-- To demonstrate the symmetry-based separation of STR, PHE, and IPHE contributions to $V_{yx}$, we have performed measurements of the electric Hall effect in an in-plane geometry on the room-temperature Weyl ferromagnet \ce{Fe3Sn}. This material is a representative of the kagome metal series \ce{X_p Y_q} (where X = Fe, Co, Ni and Y = Sn, In, Ge). \ce{Fe3Sn} exhibits an easy-plane magnetocrystalline anisotropy~\cite{prodan2023large}, resulting in a net in-plane magnetization $M_x$ along the $x$-direction with small out-of-plane canting~\cite{sankar2025room}. Therefore, $C_{3z}$ symmetry is broken by $M_x$, favoring the appearance of magnetoanisotropy effects in $V_{xy}$. It was previously shown that epitaxial thin films of \ce{Fe3Sn} exhibit the AIPHE at room temperature when ${\bf B}_\parallel$ is applied~\cite{sankar2025room}.

\begin{figure}
    \centering
    \includegraphics[width=\columnwidth]{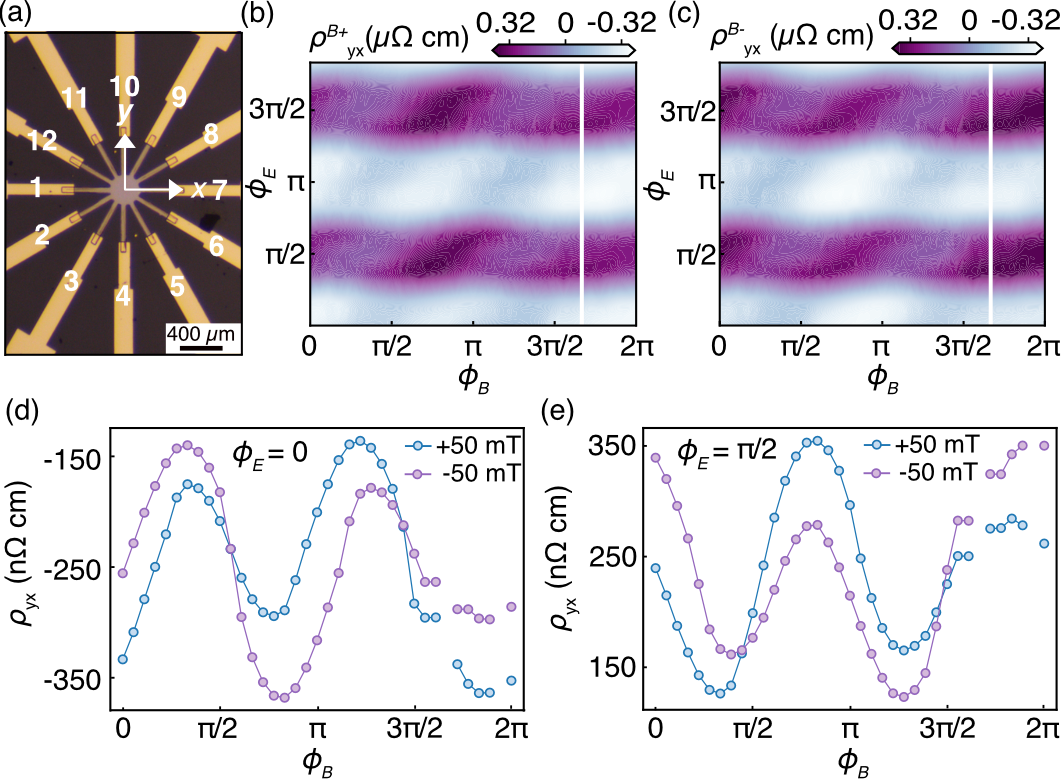}
    \caption{(a) Optical micrograph of a circular Fe\(_3\)Sn(0002)/Pt(111)/sapphire(0001) thin-film device. The image includes contact labels (numbers) and the \(x\)-\(y\) coordinate axes. The in-plane magnetization \(M_x\) is oriented along \(+x\). (b) and (c), shown are the full angle dependence of $\rho^{B+}_{yx}$ and $\rho^{B-}_{yx}$, respectively within the $(\phi_B, \phi_E)$-plane recorded on an Fe$_3$Sn device  at a temperature $T = 300\,\mathrm{K}$ and an in-plane magnetic field $B_\parallel^\pm = \pm50\,\mathrm{mT}$. (d) and (e), shown is the transverse electric resistivity $\rho_{yx}$ as a function of $\phi_B$ recorded at $\phi_E=0$ and $\phi_E=\pi/2$, respectively.}
    \label{fig:3}
\end{figure}

A $d=\SI{30}{nm}$ thick epitaxial film of \ce{Fe3Sn} was grown via molecular beam epitaxy (MBE) on c-axis sapphire using a platinum buffer layer, as described elsewhere~\cite{sankar2025room}. Device fabrication employed standard Ar$^+$ ion milling and photolithography techniques to pattern a circular Hall bar structure with a \SI{200}{\micro\meter} diameter and 12 terminals, as depicted in Fig.~\ref{fig:3}(a). The device was subsequently wire-bonded to a 12-terminal puck and characterized in a Quantum Design Physical Property Measurement System (PPMS 6000).

This 12-terminal measurement configuration facilitates $\phi_E$-dependent     measurements of $V_{xy}$ with resolution $\Delta\phi_E=\pi/6$. The coordinate system is defined such that $\phi_B=\phi_E=0$ along the $x+$ direction. For example, the application of a bias current $I$ between contacts 1--7 and 4--10 corresponds to $\phi_E = 0$ and $\phi_E = \pi/2$, respectively, as shown in Fig.~\ref{fig:3}(a). Measurements as a function of $\phi_B$ are realized by rotating an externally applied magnetic field $\mathbf{B}_{\parallel} = B(\cos\phi_B,\sin\phi_B,0)$ within the sample plane using a sample rotator, that is, $\phi_E$ and $\phi_B$ can be controlled independently.

In Fig.~\ref{fig:3}, (b) and (c), we present the transverse resistivities $\rho_{yx}^{B+}(\phi_B,\,\phi_E)$ and $\rho_{yx}^{B-}(\phi_B,\,\phi_E)$ measured under an applied in-plane field $B_\parallel=\SI{50}{mT}$ and after field reversal $B_\parallel=-\SI{50}{mT}$, respectively. We calculate $\rho_{yx}^{B+,\,B-}=V_{yx}I/d$. The twelve data sets recorded at different $\phi_E$ values were interpolated along the $\phi_E$-axis to enhance visual clarity. Both data sets, shown in panels b and c, exhibit a pronounced background, which is $\pi$-periodic with respect to $\phi_E$. At the same time, additional $\phi_E$ and $\phi_B$ dependent modulations of $\rho_{yx}$ can be detected. These modulations exhibit a clear lack of symmetry with respect to $\phi_E$, $\phi_B$ as well as between positive and negative magnetic field directions, as can be seen from individual $\rho_{xy}^{B+}(\phi_B)$ and $\rho_{xy}^{B-}(\phi_B)$ traces [Fig.~\ref{fig:3},~(d) and (e)]. Consequently, these unsymmetrized raw data sets yet provide limited direct physical insights, preventing the interpretation of experimental data.

\vspace{0.25cm}
{\em Analysis and Discussion}.--
We now apply our symmetry-guided analysis and first decompose $\rho_{yx}(\phi_B,\,\phi_E)$ into magnetic field symmetric $\rho_{\mathrm{sym}}(\phi_B,\,\phi_E)$ and antisymmetric $\rho_{\mathrm{asym}}(\phi_B,\,\phi_E)$ components,
\begin{equation}
   \rho_{\text{sym}}(\phi_B, \phi_E) = \frac{\rho_{xy}(\phi_B, \phi_E, \mathbf{B}_{\parallel}^+) + \rho_{xy}(\phi_B, \phi_E, \mathbf{B}_{\parallel}^-)}{2}, 
\end{equation}
and 
\begin{equation}
    \rho_{\text{asym}}(\phi_B, \phi_E) = \frac{\rho_{xy}(\phi_B, \phi_E, \mathbf{B}_{\parallel}^+) - \rho_{xy}(\phi_B, \phi_E, \mathbf{B}_{\parallel}^-)}{2}.
\end{equation}

\begin{figure}
    \centering
    \includegraphics[width=\columnwidth]{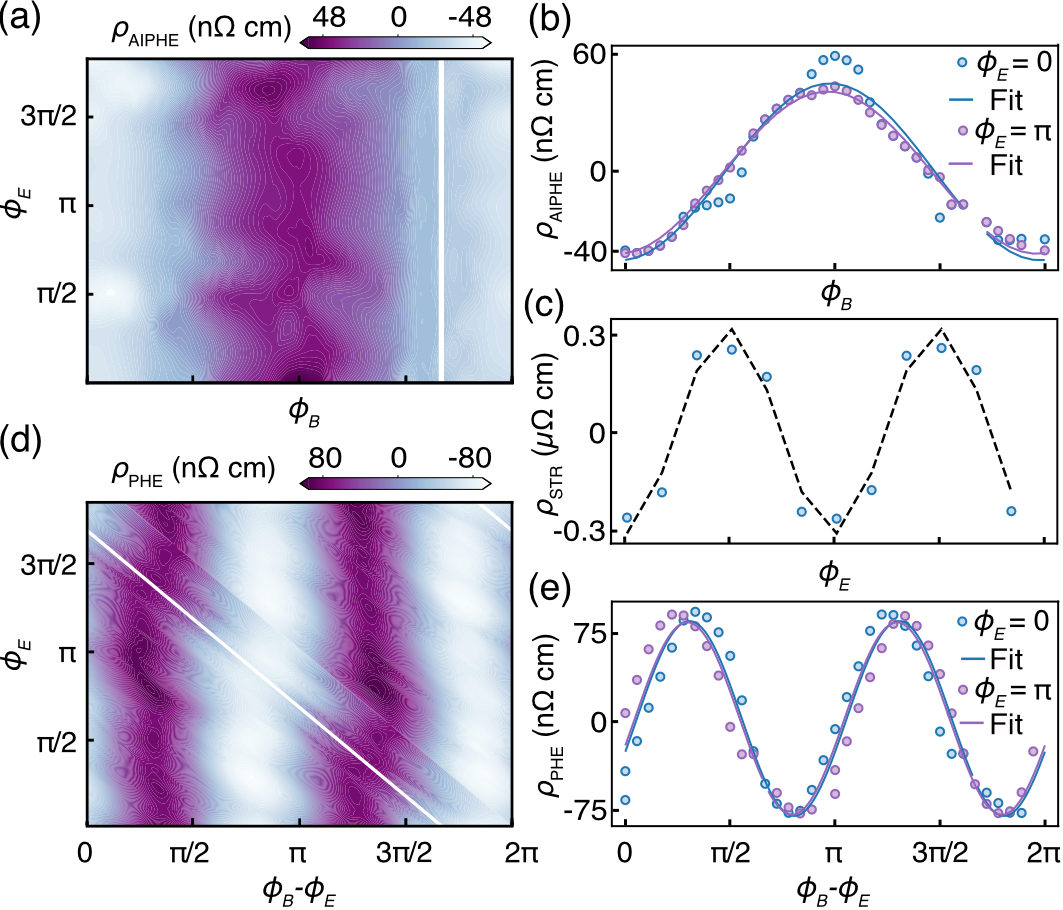}
    \caption{ (a) Shown is the full angular dependence of the anomalous in-plane Hall resistivity $\rho_\mathrm{AIPHE}$ within the ($\phi_B$, $\phi_E$)-plane ($T = 300\,\mathrm{K}$ and $B_\parallel = 50\,\mathrm{mT}$). (b) Shown is $\rho_\mathrm{AIPHE}(\phi_B)$ at $\phi_E = 0$ and $\phi_E = \pi$. The data (symbols) and fits to the data (solid lines) are shown. (c) Shown is the symmetric transverse resistivity $\rho_{\mathrm{STR}}(\phi_E)$ ($\phi_B = \pi/2$, $T = 300\,\mathrm{K}$, and $B_\parallel = 20\,\mathrm{mT}$). (d) Shown is the full angular dependence of the planar Hall effect resistivity $\rho_\mathrm{PHE}$ within the $(\phi_B - \phi_E, \phi_E)$-plane ($T = 300\,\mathrm{K}$ and $B_\parallel = 50\,\mathrm{mT}$). (e) Shown is $\rho_\mathrm{PHE}(\phi_B - \phi_E)$ at $\phi_E = 0$ and $\phi_E = \pi$. The data (symbols) and fits to the data (solid lines) are shown.}
    \label{fig:4}
\end{figure}

When neglecting contributions from the OHE and AHE (see Ref.~\cite{SI} for discussion), it follows for the $\phi_B$-dependent antisymmetric component of the resistivity tensor $\rho_{\text{asym}}=\rho_{\text{AIPHE}}$. In Fig.~\ref{fig:4}(a), we display $\rho_{\text{AIPHE}}$ as a function of $\phi_B$ and $\phi_E$. We find that $\rho_{\text{AIPHE}}$ is independent of $\phi_E$ but exhibits a sinusoidal dependence on $\phi_B$. This is also highlighted in Fig.~\ref{fig:4}(b), where we plot $\phi_B$-dependent traces of $\rho_{\text{AIPHE}}$ at different constant $\phi_E$-values ($\phi_E=0$ and $\phi_E=\pi$).

We can identify the AIPHE origin of $\rho_{\text{asym}}(\phi_B)$ via its unique cosine angle-dependence with respect to $\phi_B$. This understanding is verified by applying a least-square fit 
\[
\rho_{\text{AIPHE}}(\phi_B) = \rho_0 + \rho_{\text{AIPHE}}^0 \cos(\phi_B + \delta)
\]
which accurately captures the detailed angle-dependence. The phase offset $\delta$ accounts for minor systematic angular deviations arising from the imperfect alignment of $\phi_B$ and $\phi_E$ with respect to the magnetization axis $M_x$ during measurement and $\rho_0$ accounts for a small vertical offset, such as induced by AHE and OHE. 

Quantitatively, the analysis yields $|\rho_{\text{AIPHE}}| = 44.0 \pm 1.9$ n$\Omega\cdot$cm at $\phi_E = 0$ and $|\rho_{\text{AIPHE}}| = 40.4 \pm 0.7$ n$\Omega\cdot$cm at $\phi_E = \pi$, demonstrating excellent consistency between the two measurement configurations. The corresponding phase offset $\delta$ remains below $6^\circ$, indicating minimal angular misalignment and confirming the robustness of the fit. The complete set of fit parameters is listed in the Supplementary Material \cite{SI}.

To separate the contributions from STR and PHE to $\rho_{\text{sym}}(\phi_E, \phi_B)$, we decompose $\rho_{\text{sym}}(\phi_E, \phi_B)$ into a $\phi_E$-dependent part, $\rho_{\text{STR}}(\phi_E)$, and a $\phi_B-\phi_E$-dependent part, $\rho_{\text{PHE}}(\phi_B - \phi_E, \phi_E)$. The full $\rho_{\text{sym}}(\phi_E, \phi_B)$-map is shown in the Supplementary Materials~\cite{SI}.

In Fig.~\ref{fig:4}(c), we plot the resulting $\rho_{\text{STR}}(\phi_E)$, which only depends on the relative angle $\phi_E$ between the bias current direction (parallel to ${\bf E}$) and $M_x$ (parallel to $+x$-direction). Performing a least-square fit using $\rho_{\text{STR}}(\phi_E) = \rho_{STR}^0\sin (2\phi_E + \phi')$ confirms the $\pi$-periodic sinusoidal relation of $\rho_{\text{STR}}(\phi_E)$ with respect to $\phi_E$ (fit parameters are listed in Ref.~\cite{SI}). Note that $\phi'$ accounts for an angle offset between the current bias direction at $\phi_E=0$, as defined in Fig.~\ref{fig:3}(a), and ${\bf M}_{\parallel}$. The presence of STR is indeed expected, given the breaking of $C_{3z}$ symmetry by the in-plane magnetization of Fe$_3$Sn. Yet, the identification and extraction of its contribution to $\rho_{yx}$ is made possible by using a circular Hall bar with the symmetry-guided analysis presented in this study.

Finally, in Fig.~\ref{fig:4}(d), we plot $\rho_{\text{PHE}}(\phi_B - \phi_E,\,\phi_E)$ as the $\phi_B-\phi_E$-dependent contribution to $\rho_{\text{sym}}(\phi_E,\,\phi_B)$. It is apparent from the data that $\rho_{\text{PHE}}(\phi_B - \phi_E,\,\phi_E)$ exhibits a $\pi$-periodic sinusoidal dependence on the relative angle between ${\bf B}_\parallel$ and ${\bf E}$. To quantitatively analyze this angular dependence, we have extracted two representative traces at fixed $\phi_E = 0$ and $\phi_E = \pi$, shown  in Fig.~\ref{fig:4}(e), and performed least-square fits using:
\[
\rho_{\text{PHE}}(\phi_B - \phi_E) = \rho_0^{\text{PHE}} \sin\left[2(\phi_B - \phi_E)\right]
\]
where $\rho_0^{\text{PHE}}$ is the amplitude of the planar Hall resistivity ($|\rho_{\text{PHE}}| = 83.7 \pm 0.4$ n$\Omega\cdot$cm at $\phi_E = 0$ and $|\rho_{\text{PHE}}| = 82.6 \pm 0.4$ n$\Omega\cdot$cm for $\phi_E = \pi$). The results of this analysis confirm the excellent agreement between the experimental data and the theoretically expected dependence of $\rho_{\text{PHE}}$ on $\phi_B - \phi_E$~\cite{zhong2023recent}. While we can detect small variation of $\rho_{\text{PHE}}$ with respect to $\phi_E$, probably owing to the accuracy of our sample rotator probe, overall, $\rho_{\text{PHE}}$ is independent of $\phi_E$. This observation is consistent with nature of the PHE as a magnetic-field induced anisotropic magnetoresistance effect.

\vspace{0.25cm}
{\em Conclusion}.--Our work establishes a general framework for distinguishing and separating in-plane Hall responses. Leveraging a circular 12-terminal Hall bar geometry, we show that rotations of the electric and in‑plane magnetic field can serve as effective control knobs for topological transport signals. We showcase this methodology by analyzing the transverse electric voltage response of \ce{Fe3Sn} under an in-plane measurement geometry. While the combined contributions of the symmetric transverse resistance, the planar Hall effect, and the anomalous in-plane Hall effect lead to irregular characteristics of the transverse electric resistivity signal [Fig.~\ref{fig:3}, (d) and (e)], the application of symmetry-guided symmetrization and data analysis facilitates their separation and quantitative analysis. This methodology will guide future studies of topological and magnetic quantum materials using in-plane measurement geometries and opens new opportunities for exploiting in‑plane Hall effects in functional devices, including magnetic field sensors and spin‑orbitronics.

\vspace{0.25cm}
{\em Acknowledgement}.--The authors appreciate valuable discussions with Xi Dai. This work was primarily supported by the Hong Kong Research Grant Council (Grant Nos.\,26304221, 16302422, 16302624, 16304525, and C6033-22G awarded to BJ), the Croucher Foundation (Grant No.\,CIA22SC02 awarded to BJ), and the National Key R$\&$D Program of China (Grant No.\,2021YFA1401500 awarded to JL). JL further acknowledges support from the Hong Kong Research Grants Council (Grant Nos.\,16306722, 16304523, and C6046-24G).

\clearpage
\section{References}
\bibliographystyle{apsrev4-1}   
\bibliography{Anisotropy}       

\end{document}